\documentclass[aps, prl, twocolumn, superscriptaddress, noshowpacs, floatfix]{revtex4-2}

\usepackage{amsmath, amssymb, mathtools, bm}
\usepackage{graphicx}
\usepackage{color}
\usepackage{hyperref}
\usepackage{ulem}

\begin{document}

\title{Linear Mode Conversion in Ultramagnetized Pair Plasmas: Single-Parameter Scaling}

\author{Dawei Dai}
\affiliation{Department of Physics \& Astronomy, Wilder Laboratory, Dartmouth College, Hanover, NH 03755, USA}

\author{Ashley Bransgrove}
\affiliation{Princeton Center for Theoretical Science, Princeton University, Princeton, NJ 08544, USA}
\affiliation{Department of Astrophysical Sciences, Princeton University, Princeton, NJ 08544, USA}

\author{Anirudh Prabhu}
\affiliation{Princeton Center for Theoretical Science, Princeton University, Princeton, NJ 08544, USA}

\author{Jens F. Mahlmann}
\affiliation{Department of Physics \& Astronomy, Wilder Laboratory, Dartmouth College, Hanover, NH 03755, USA}
\begin{abstract}
In neutron star (NS) magnetospheres, plasma waves propagate as normal modes with distinct propagation dynamics that strongly influence observable signals. This letter presents a unified theory of linear mode conversion between Alfvén (A), superluminal ordinary (O), and extraordinary (X) modes, incorporating the effect of magnetic-field geometry and local plasma response. Magnetic field‑line curvature induces A--X conversion for low frequencies and O--X conversion at high frequencies, whereas plasma gradients alone do not drive X-mode coupling. We show that a single dimensionless parameter controls both conversion channels. The conversion efficiency follows the universal nonadiabatic transition probability of a multilevel quantum system. Efficient conversion occurs within a narrow angular window between the wave vector and magnetic field, localizing potential conversion sites in the NS magnetosphere. This linear mechanism naturally accounts for complex polarization features observed in pulsars and some fast radio bursts.
\end{abstract}
\maketitle

Astrophysical radio transients are brief, intense bursts of radio waves originating from some of the universe's most extreme environments. Pulsed radio emission from neutron stars \cite[NSs,][]{Philippov2022} and fast radio bursts \cite[FRBs,][]{CHIME2023} both probe coherent processes in strongly magnetized plasmas.
After emission, propagation through the surrounding plasma can significantly alter the outgoing signal. Interpreting radio transients therefore requires understanding the microphysics of wave propagation.

NS magnetospheres contain strongly magnetized, relativistic electron-positron plasmas supporting three normal wave modes \cite{1986ApJ...302..120A}: extraordinary (X) modes, subluminal (Alfvén/A) and superluminal ordinary (O) modes. Each mode has different generation mechanism and propagation behavior. Alfvén waves can be generated by plasma instabilities
\cite{Beloborodov2013a,melrose2017coherent,zeng2025} but likely cannot escape from the magnetosphere due to severe Landau damping \cite{1986ApJ...302..120A}. Non-steady plasma discharges can generate O-modes with superluminal phase speeds that can escape from the host magnetosphere \cite{PhysRevLett.124.245101}. X-modes do not couple to plasma processes, they are produced by a limited set of emission mechanisms \cite{Philippov2019,Plotnikov2019} and propagate as vacuum electromagnetic (EM) waves in the magnetosphere. 

Radio polarimetry of pulsars and magnetars suggests the presence of orthogonal modes in escaping signals \cite{manchester1975observations,cordes1978orthogonal}. Since Alfvén waves cannot escape the magnetosphere and direct X‑mode generation is inefficient, mode conversion during propagation is required to account for the observed radio properties. Existing works treat individual conversion channels or specific plasma conditions separately, only qualitatively predicting most efficiencies. \citet{Petrova2001} considered linear O--X conversion due to field line bending when wave vector and magnetic field lines are nearly aligned due to geometry and refraction. 
Several works considered A--X conversion in force-free electrodynamics \cite{Yuan_2021,Mahlmann2024,Bernardi2025,Chen2025} and curvature radiation of charged bunches moving along curved field lines \cite{Gil_2004}. A--O conversion due to plasma density gradient effects is inefficient for pulsar conditions \cite{1996AstL...22..482B}. Until now, no unified framework has quantified conversion efficiency among all three magnetospheric modes in a general way.

This Letter presents a unified framework that describes conversion between A-, O-, and X-modes driven by field line bending and plasma gradient effects. We show that plasma mode conversion is analogous to the non-adiabatic transition phenomena in multilevel quantum systems, with a single dimensionless parameter determining the conversion efficiency. This framework can explain circular polarization, position angle (PA) jumps and depolarization effects in radio pulsars and FRBs observations.
\begin{figure}
    \centering
    \includegraphics[width=0.95\linewidth]{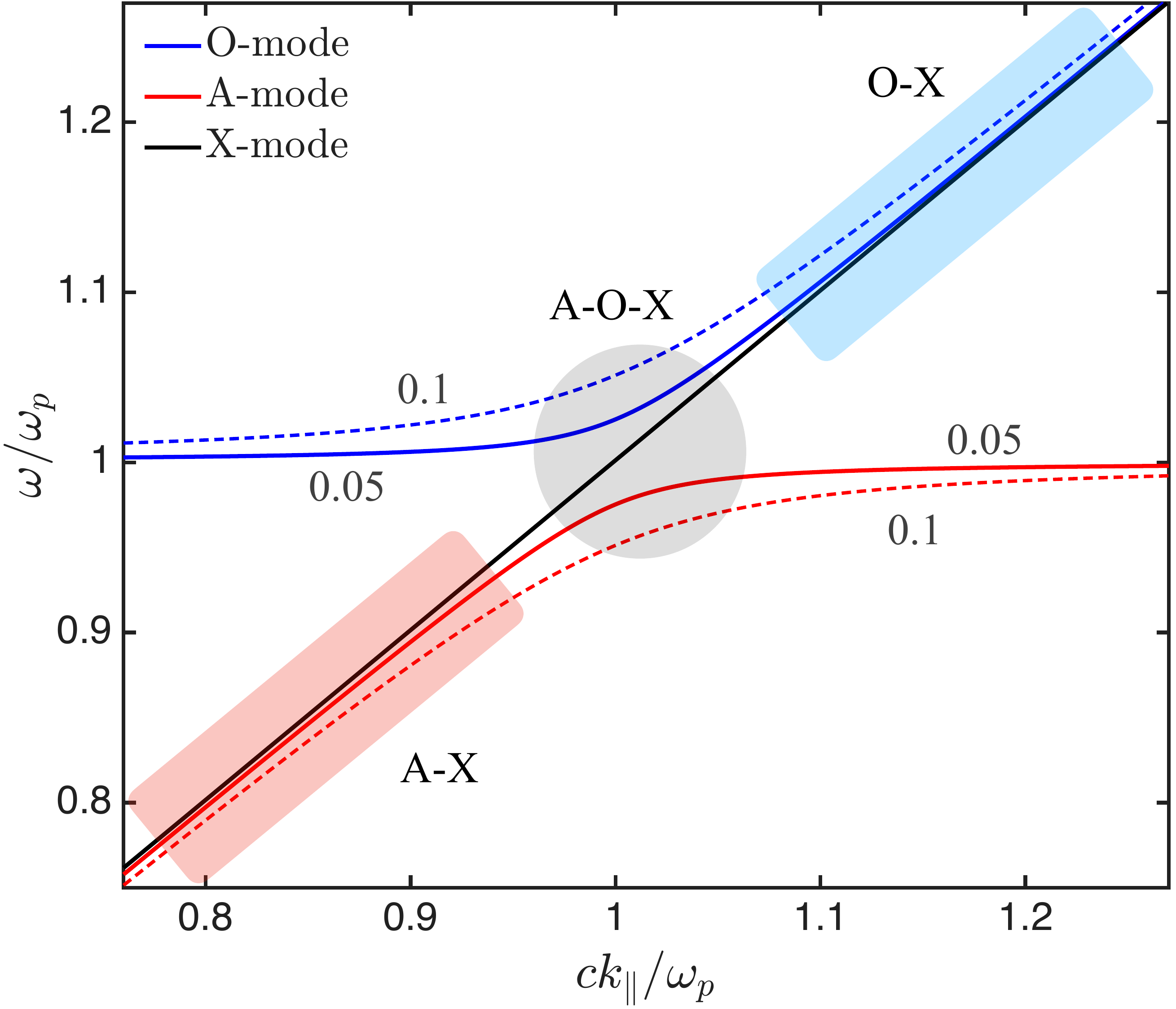}
    \caption{Cold plasma dispersion curves of A- (red), O- (blue) and X-modes (black) for $\gamma_0=1$. Solid lines are dispersion curves with $k_{\perp}/\omega_\mathrm{p}=0.05$, dashed lines with $k_{\perp}/\omega_\mathrm{p}=0.1$. The red shaded region corresponds to A--X conversion, the blue shaded region to O--X coupling, the gray shaded region to A--O--X conversion.}
    \label{fig:dispersion curve} 
\end{figure}
In NS magnetospheres, waves follow the dispersion relation for ultramagnetized pair plasmas (see below):
\begin{align}
	\left(\omega^2-c^2 k^2\right)\left[\left(\omega^2-c^2 k_{\|}^2\right)\left(1-\frac{\omega_{\text{p,eff}}^2}{\omega^2}\right)-c^2 k_{\perp}^2\right]=0\,.
 \label{Dispersion relation}
\end{align}
Here,  $c$ is the speed of light, $\omega$ is the wave frequency, and $k=|\boldsymbol{k}|$ is the wave vector, projected along ($\parallel$) and perpendicular ($\perp$) to the background magnetic field. The effective plasma frequency is $\omega_{\text{p,eff}}=\omega_\mathrm{p}/[\gamma_0^{3/2}\left(1-\beta_0 c k_{\|}/\omega\right)]$, with $\omega_\mathrm{p}^2=4\pi n_\mathrm{e}e^2/m_\mathrm{e}$, electron density $n_\mathrm{e}$, and electron mass $m_\mathrm{e}$. We assume a cold-plasma streaming along $\boldsymbol{B}$ with velocity $\beta_0$, $\gamma_0=(1-\beta_0^2)^{-1/2}$ is the Lorentz factor. Figure~\ref{fig:dispersion curve} shows dispersion curves for the three solution branches of Equation~(\ref{Dispersion relation}), corresponding to the three eigenmodes of plasma waves. The resonance frequency
\begin{align}
\omega_{\rm{res}}=\omega_{\text{p,eff}}\left(\omega=ck_\parallel\right)=\frac{\omega_\mathrm{p}}{\gamma_0^{3/2}(1-\beta_0)}\,
\label{mode crossing}
\end{align} 
marks an avoided crossing of the ordinary mode dispersion branches. Small $k_{\perp}$ brings dispersion curves close to each other, so modes are almost indistinguishable and efficient conversion can be realized. We identify three relevant mode-conversion regimes:  A--X conversion below the resonance frequency (red shaded), O--X conversion above the resonance frequency (blue shaded), and A--O--X conversion at the resonance frequency (gray shaded). Conversion between all three modes is restricted to a narrow region around $\omega\approx\omega_{\rm res}$, and we discuss it separately (see below).

For waves in an inhomogeneous plasma with a time-independent dispersion relation, the frequency is constant while the wave vector may vary. Since X-modes propagate as EM modes that are not coupled to the background plasma, plasma gradient effects only contribute to A--O conversion near the avoided crossing of the dispersion branches. In the following, we only consider field line curvature-driven conversion away from the resonance frequency, and show that both A--X and O--X conversion channels can be described using the same formula.

To formulate a linear theory of mode conversion, we 
assume $k_{\perp}/k_{\|}\ll1$, as significant conversion only happens between branches with close-by dispersion curves. Then, the parallel refractive index is $n_{\|}=ck_{\|}/\omega\approx 1+\mathcal{O}(k_{\perp}^2)$, and $\omega_{\text{p,eff}}\approx\omega_{\rm res}$. We justify this assumption below for realistic parameters. We also assume the system length scale $L$ is small compared with the inhomogeneity length scale given by the curvature radius $\rho$, thus, the wave equation can be linearized. This assumption is justified separately (see below). 

\begin{figure}
    \centering
    \includegraphics[width=0.9\linewidth]{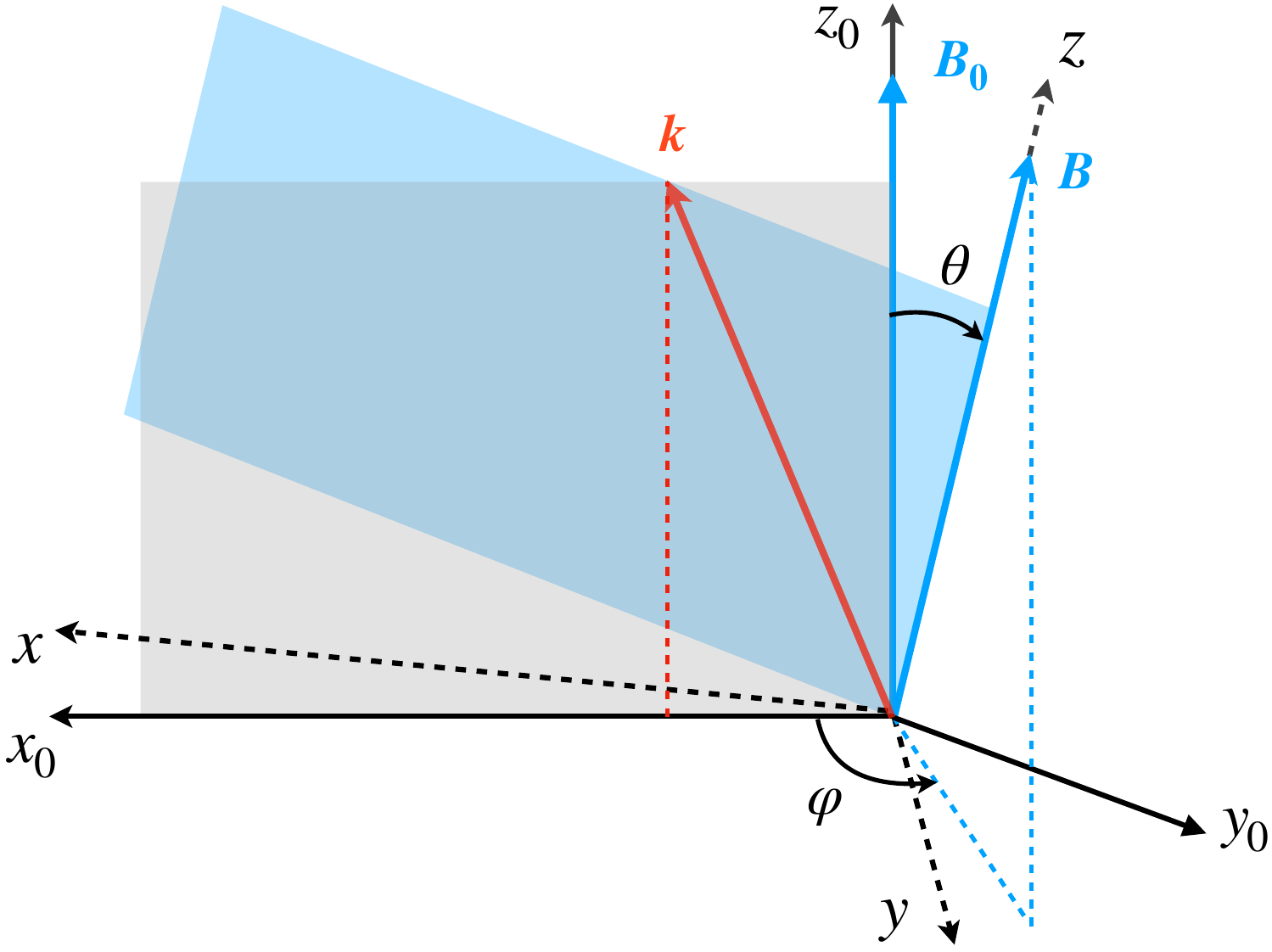}
    \caption{Illustration of the coordinate system adjusting to field line bending. Shaded surfaces highlight the $\boldsymbol{k}-\boldsymbol{B}$ plane before (grey) and after (blue) the rotation. As the wave propagates from $z_0$ to $z$, the magnetic field vector rotates by angles $\theta$ with the $\boldsymbol{z_0}$-axis, and $\varphi$ in the $\boldsymbol{x_0}-\boldsymbol{y_0}$ plane. The local coordinate system $(\boldsymbol{x},\boldsymbol{y},\boldsymbol{z})$ has $\boldsymbol{z}\|\boldsymbol{B}$, and $\boldsymbol{x}$ in the $\boldsymbol{x_0}-\boldsymbol{y_0}$ plane.}
    \label{fig:illustration} 
\end{figure}
\begin{figure*}
\includegraphics[width=1\textwidth]{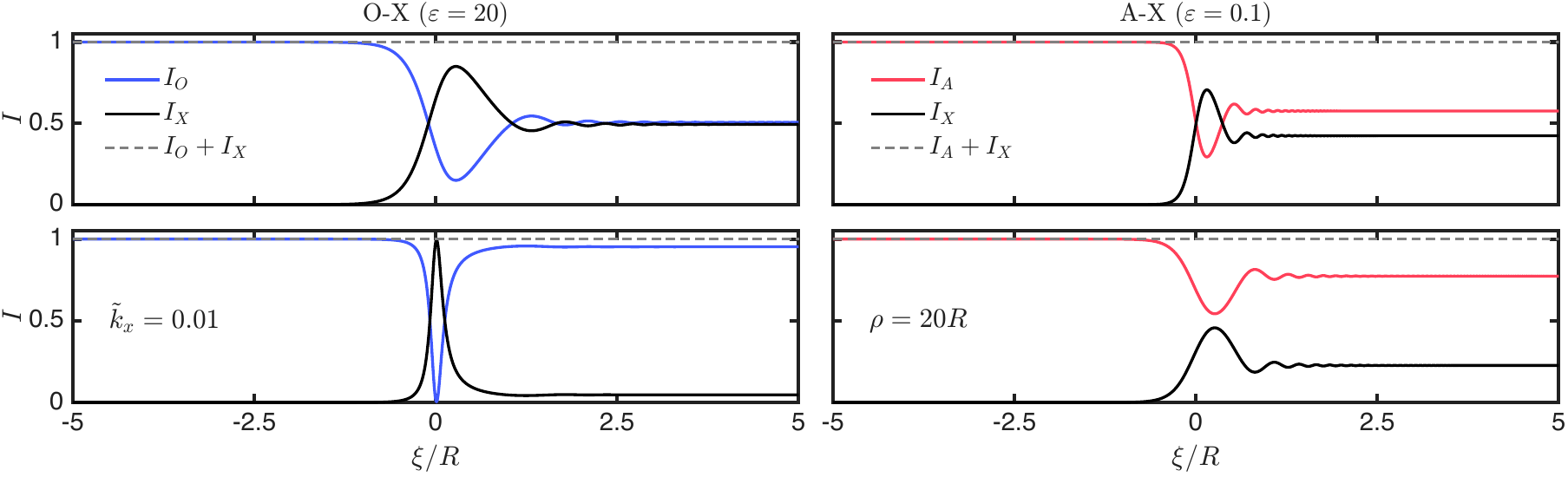}
\vspace{-16pt}
\caption{Mode energy evolution during O--X conversion ($\varepsilon=20$, left panels) and A--X conversion ($\varepsilon=0.1$, right panels). Cases with $k_{\rm res}=1$, $\tilde{k}_x=0.05$, $\rho=10R$ ($R=10\rm{km}$), and $\varphi=\pi/2$ are shown in the top row. In the bottom row, one parameter varies as indicated while the others remain fixed. Normal modes can only be defined and distinguished outside the conversion region. Nevertheless, for illustration we track the mode electric fields and energy in this figure and Equation~(\ref{basis}).}
\label{fig:Evolution}
\end{figure*} 

In NS magnetospheres, the plasma dielectric tensor $\boldsymbol{\epsilon}$ is commonly defined in local Cartesian coordinates with the $\boldsymbol{z}$ direction along the background magnetic field. These local coordinate systems rotate with the bending field line along the wave path, and the wave equation with $\boldsymbol{\epsilon}$ defined in a single local frame cannot capture the global wave evolution. To construct a global wave equation, we transform $\boldsymbol{\epsilon}$ from the local frames $(\boldsymbol{x},\boldsymbol{y},\boldsymbol{z})$ with $\boldsymbol{z}\|\boldsymbol{B}$ to a fixed coordinate system $(\boldsymbol{x_0},\boldsymbol{y_0},\boldsymbol{z_0})$ defined at a reference point $z=z_0$. As illustrated in Figure~\ref{fig:illustration}, the background magnetic field $\boldsymbol{B_0}$ rotates to a new orientation $\boldsymbol{B}$, characterized by an angle $\theta$ with the $\boldsymbol{z_0}$-axis and an azimuthal angle $\varphi$ in the $\boldsymbol{x_0}-\boldsymbol{y_0}$ plane. The $\boldsymbol{x}$-axis is in the $\boldsymbol{x_0}-\boldsymbol{y_0}$ plane. The global wave equation in the fixed coordinate system then reads
\begin{align}
    \boldsymbol{k} \times(\boldsymbol{k} \times \boldsymbol{E})+\frac{\omega^2}{c^2}\mathbf{T}^{-1}\boldsymbol{\epsilon} \mathbf{T}\cdot \boldsymbol{E}=\boldsymbol{0}\,,
    \label{wave equation vector}
\end{align}
where $\mathbf{T}$ is the transformation matrix between the local and the fixed coordinate system,
\begin{equation}
\mathbf{T} = \begin{pmatrix}
    \sin\varphi & -\cos\varphi & 0 \\
    \cos\theta\cos\varphi & \cos\theta\sin\varphi & -\sin\theta \\
    \sin\theta\cos\varphi & \sin\varphi\sin\theta & \cos\theta
\end{pmatrix}.
\label{transformation matrix}
\end{equation}

We analyze wave behavior close to the coupling point, in a linear expansion around $z_0$:
\begin{align}
        z=z_0+\xi,\ \ \ \theta(z)=\frac{\xi}{\rho}, \ \ \ k_z=k_0\left(1+\lambda\right)\,,
\label{linear expansion}
\end{align}
where $\theta\ll1, \lambda\ll1, k_0=\omega/c$. For simplicity, we do not include twisting of magnetic field lines and assume $\varphi$ to be constant. Using the linearized Equation~(\ref{linear expansion}), and substituting Equations~(\ref{dieletric tensor}) and~(\ref{transformation matrix}) into (\ref{wave equation vector}), we find
\begin{equation}
\begin{pmatrix}
    2 \lambda+\delta_x^2 & \delta_x \delta_y & -\tilde{k}_x+\delta_x / \varepsilon \\
    \delta_x \delta_y & 2 \lambda+\tilde{k}_x^2+\delta_y^2 & \delta_y / \varepsilon \\
    -\tilde{k}_x+\delta_x / \varepsilon & \delta_y / \varepsilon & -1 + \varepsilon^{-2}
\end{pmatrix} \cdot \bm{E} = 0.
\label{expanded matrix}
\end{equation}
Here, we used the normalizations $\tilde{k}_x=k_x/k_0$, $\varepsilon = \omega/\omega_{\rm res}$, $\delta_x=\theta\cos\varphi/\varepsilon$, and $\delta_y=\theta\sin\varphi/\varepsilon$. Each element of the matrix is retained to leading order. Because $E_z$ is an order smaller than $E_x$ and $E_y$, the upper-left $2\times2$ block is preserved to second order. The two eigenvalues of Equation~(\ref{expanded matrix}) capture the independent evolution of O/A and X-mode wave vectors
\begin{align}
    &\lambda^{\rm{O/A}}=\frac{\delta_x^2+\delta_y^2+\tilde{k}_x^2-2\delta_x \tilde{k}_x/\varepsilon}{2(-1+\varepsilon^{-2})},\qquad \lambda^{\rm X}=-\frac{\tilde{k}_x^2}{2}\,.
\label{eigenvalue}
\end{align}
This expression applies to both A-modes ($\epsilon<1$) and O-modes ($\epsilon>1$) in their respective frequency range. To calculate the electric field evolution, we recall $\lambda=k_z/k_0-1$ and substitute $k_z\approx-i{\rm d}/{\rm d}\xi$ to obtain the coupled differential equations
\begin{align}
\left(i\frac{{\rm d}}{{\rm d}\xi}+k_0\right)
\begin{pmatrix} E_x \\ E_y \end{pmatrix}
= \frac{k_0}{2(1-\varepsilon^{-2})}
\begin{pmatrix}
H_{xx} & H_{xy} \\
H_{yx} & H_{yy}
\end{pmatrix}\cdot
\begin{pmatrix} E_x \\ E_y \end{pmatrix}\,,
\label{ODE}
\end{align}
where $H_{xx}=\delta_x^2+\tilde{k}_x^2-2\tilde{k}_x\delta_x/\varepsilon, \ H_{yy}=\tilde{k}_x^2+\delta_y^2-\tilde{k}_x^2/\varepsilon^2$ and $H_{xy}=H_{yx}=\delta_y(\delta_x-\tilde{k}_x/\varepsilon)$. The electric field evolution $\boldsymbol{E}=\boldsymbol{E}(\xi)$ can then be determined for given initial conditions. Field amplitudes of different modes are obtained by an adiabatic basis projection:
\begin{align}
\begin{split}
\boldsymbol{E}=E^{\rm X}&\boldsymbol{e}_{\rm X}+E^{\rm O/A}\boldsymbol{e}_{\rm O/A}\,,\\
\boldsymbol{e}_{\rm X}\propto(\delta_y,\tilde{k}_x/\varepsilon-\delta_x)&,\quad
\boldsymbol{e}_{\rm O/A}\propto(\tilde{k}_x/\varepsilon-\delta_x,-\delta_y)\,.
\end{split}
\label{basis}
\end{align}
 Here, $\boldsymbol{e}_{\rm X}$ and $\boldsymbol{e}_{\rm O/A}$ are the orthogonal eigenvectors of each mode that follow the rotation of the $\boldsymbol{k}-\boldsymbol{B}$ plane with corresponding eigenvalues $\lambda_{\rm X}$ and $\lambda_{\rm O/A}$. Ordinary (O/A) modes share the same polarization and eigenvector in their respective frequency range. We omit $E_z$ as it is an order of magnitude smaller than $E_{x,y}$, thus all modes are polarized in the $\boldsymbol{x_0}-\boldsymbol{y_0}$ plane. 

We integrate Equation~(\ref{ODE}) numerically using a stepwise matrix exponential propagator in \textsc{Matlab}. The energy flux of each mode is proportional to the squared electric field amplitude. Figure~\ref{fig:Evolution} shows examples of wave evolution during A--X and O--X conversion for different parameters. We inject pure A- or O-modes from the left boundary, and the conversion efficiency is determined from the outgoing X-mode energy flux at the right boundary. Although Equation~(\ref{ODE}) has no analytic solution, there is a simple asymptotic estimate for the total conversion efficiency:
\begin{align}
    \eta_{\rm{O/A\rightarrow X}}=\frac{I_{\rm{X}}(+\infty)}{I_{\rm{O/A}}(-\infty)}=\frac{|E^{\rm X}(+\infty)|^2}{|E^{\rm O/A}(-\infty)|^2}\,.
\end{align}
The asymptotic approximation is valid when the length scale of the conversion region $l$ is much smaller than the system size $L$, this assumption is justified separately (see below). Asymptotic solutions of the coupled first-order ODEs in Equation~(\ref{ODE}) are extensively studied in the context of non-adiabatic transitions in quantum physics. Following \citet{landau1932theory} and \citet{zener1932non} for calculating state transition probabilities, we 
estimate the conversion efficiency $\eta$ as
\begin{align}
\begin{split}
        &\eta_{\rm{O/A\rightarrow X}}=\sqrt{2}p(1-p)^{\alpha},\quad\alpha=0.6\,\\&p=\exp(-\Delta),\quad\Delta=\frac{2\tilde{k}_x^3k_0\rho\sin^3\varphi}{3|\varepsilon^2-1|}\,.
\end{split}
\label{fit}
\end{align}
where $\tilde{k}_x$ is the normalized wave vector in the $\boldsymbol{x_0}$-direction, and $\varepsilon=\omega/\omega_{\rm res}$ is the frequency parameter. This formula applies to both A--X and O--X conversion. Its functional form resembles the $2p(1-p)$ formula proposed by \citet[ZN,][]{nakamura2012nonadiabatic}, and reduces to an exponential decay $\eta\rightarrow\sqrt{2}p$ in the adiabatic limit ($\Delta\gg1$), reproducing the state transition probability outlined by \citet[DDP,][see below]{davis1975nonadiabatic}.  $\Delta$ effectively encodes the ratio of the refractive index difference between O/A and X-modes ($n_{\rm X}-n_{\rm O/A}\sim \tilde{k}_x^2/|\varepsilon^2-1|$), and the rotation velocity of the $\boldsymbol{k}-\boldsymbol{B}$ plane, $1/k_x\rho$. This is a direct analog of the Landau–Zener (LZ) transition probability, determined by the ratio of the avoided-crossing gap to the sweep rate.

\begin{figure}
    \centering
    \includegraphics[width=0.95\linewidth]{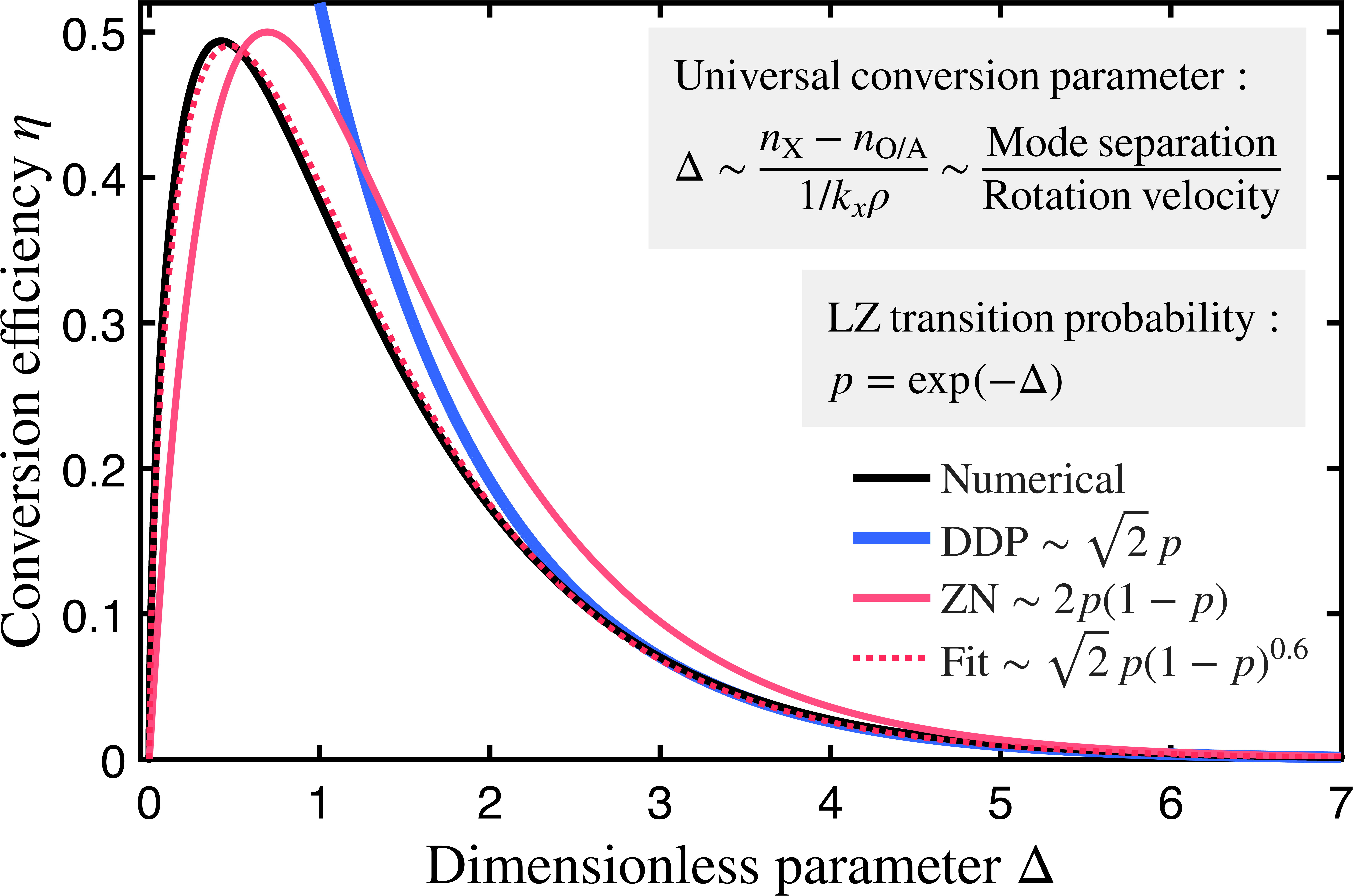}
    \caption{A--X and O--X conversion efficiencies $\eta$ as a function of $\Delta$. We compare numerically obtained efficiencies (black line) with the formula proposed by \citet[][solid red line]{nakamura2012nonadiabatic} and \citet[][solid blue line]{davis1975nonadiabatic}, and the fit formulae (dashed red line, Equation~\ref{fit}). Our analysis scans a wide parameter range of azimuthal field line bending $\varphi\in[0,2\pi]$, field line curvature $\rho/R\in[1,100]$,  frequency parameter $\varepsilon\in[0.01,100]$, and normalized wave vector in the $\boldsymbol{x_0}$-direction $\tilde{k}_x\in[0.005,0.5]$. }
    \label{fig:calibration} 
\end{figure}

In Figure~\ref{fig:calibration} we vary all variables ($\tilde{k}_x$, $\rho$, $\epsilon$, $\varphi$) simultaneously and compute the conversion efficiency as a function of $\Delta$. Over a broad multidimensional parameter space, all results collapse to the a single curve (black line), showing that $\Delta$ is the only parameter determining the conversion efficiency. As shown in Equation~\ref{fit},  The  introduced $\alpha=0.6$ coefficient accurately fits the whole range of $\Delta$, and predicts a $\eta\sim \Delta^{0.6}$ in the limit $\Delta\ll1$. The peak conversion efficiency is below $0.5$ and occurs at $\Delta\sim 1$, which can be expressed as a constraint on $\tilde{k}_x$:
\begin{align}
    \tilde{k}_x\sim\left(\frac{\max(\varepsilon^2,1)}{k_0\rho}\right)^{1/3}
\label{condition}
\end{align}
The resulting cubic dependence on $\tilde{k}_x$ indicates a strong preference for quasi-longitudinal propagation for efficient conversion. For A--O conversion $(\varepsilon<1)$, $\Delta$ is proportional to the wave frequency, while O--X conversion $(\varepsilon>1)$, $\Delta$ is inversely proportional to the wave frequency. Thus, efficient conversion occurs for low frequency A- and for high frequency O-modes. O--X conversion additionally prefers low plasma densities (small $\omega_{\rm res}$). Conversion is inefficient for $\Delta\rightarrow0$, which is consistent with conversion being absent in vacuum $(\epsilon\to \infty)$ or for field line bending in the $\boldsymbol{k}-\boldsymbol{B}$ plane $(\varphi=0)$.

After conversion, the incident and converted modes co-exist, they become distinguishable and decouple from each other as the field line bending continues. Then, they propagate independently as two orthogonal linearly polarized waves before reaching the polarization limiting region, where both wave modes become elliptically polarized. If conversion is efficient, the outgoing signals are an ensemble of coherent and incoherent components of orthogonally polarized waves. The superposition of these waves may be observed as linear and circular polarization, with PA jumps and depolarization effects.

This Letter introduces a unified theory of linear wave conversion in ultramagnetized relativistic electron-positron pair plasmas. A- and O-modes with frequencies above or below the local resonance frequency (Equation~\ref{mode crossing}) can convert to X-modes when propagating along curved field lines. By comparison with time-dependent multilevel quantum systems, we derive asymptotic estimates of the conversion efficiency that are consistent with numerical results (Figure~\ref{fig:calibration}).

These results have direct implications for NS magnetospheres, where the plasma density generally decreases with radius. For waves propagating away from the star, the ratio $\omega/\omega_{\rm res}$ therefore increases along their path. This implies that O-modes may eventually couple to X-modes when the plasma density has decreased enough, while any A-mode conversion would have to occur in the inner magnetosphere where the resonance frequency is larger than the wave frequency. Using Equation~(\ref{condition}), we estimate $\tilde{k}_x$ (ie. $k_\perp/k_\parallel$ for efficient conversion) for realistic pulsar parameters:
\begin{equation}
\begin{aligned}
\tilde{k}_x^{\rm{A-X}}
&\sim 1.7\times10^{-2}
\left(r^{-1} f_9^{-1}\right)^{1/3}\qquad\quad\ \ \ \ \ \ (\omega<\omega_{\mathrm{res}}), \\
\tilde{k}_x^{\rm{O-X}}
&\sim 2.8\times10^{-4}
\left( r^2f_9 P_1
\gamma_{10}^{-1} B_{12}^{-1}
\kappa_3^{-1}\right)^{1/3}(\omega>\omega_{\mathrm{res}}).
\label{eqn:estimate}
\end{aligned}
\end{equation}
Here, $f_9$ is the wave frequency normalized to 1 GHz, $\kappa_3$ is the plasma multiplicity normalized to $10^3, r$ is the distance from the NS in NS radii, $P$ is the rotation period of NS in seconds, $\gamma_{10}=\gamma / 10$, and $B_{12}$ is the NS surface magnetic field strength normalized to $10^{12} \mathrm{G}$. We see that in most cases, our assumption of $\tilde{k}_x \ll 1$ holds. We use Equation~\eqref{eqn:estimate} to identify three potential magnetospheric conversion sites. First, the radio emission region in the inner magnetosphere. Various emission mechanisms predict wave spectra peaked at $k_{\perp}=0$, and relativistic streaming further reduces $k_{\perp}$ in the pulsar frame \cite{Gil_2004,zeng2025}. Such waves gradually become transverse as they propagate outward, therefore, the requirement for small $k_{\perp}$ favors conversion taking place near the emission site. However, in the inner magnetosphere the plasma frequency exceeds the radio wave frequency allowing only A--X conversion. Second, in localized regions, where refraction of O-modes leads to an alignment of wave vector and field lines \cite{Petrova2001} or regions with small plasma density \cite[e.g., outer gaps,][]{bransgrove2023radio}. Third, in the outer magnetosphere, where the plasma density has dropped substantially.

The presented linear mode conversion theory is based on the same mode-coupling mechanism as the polarization limiting effect \cite[PLE,][]{lyubarskii1998,petrova2000propagation,beskin2012mean}. Mode coupling occurs when the scale lengths for beats between two modes are comparable with the scale length of the inhomogeneous background, and the wave electric field can no longer adiabatically follow the orientation of the rotating $\boldsymbol{k}-\boldsymbol{B}$ plane \cite{cheng1979theory}. In the small $k_{\perp}$ limit, O/A and X modes are nearly indistinguishable, thus, coupling is enhanced. However, it simultaneously implies $n_{\|}\approx1$, which amplifies $\omega_{\text{p,eff}}$ through relativistic effects and reduces $\varepsilon$, thereby suppressing O--X conversion in relativistic plasmas. While the PLE inevitably occurs at a finite distance from the NS, linear mode conversion is not guaranteed. In this environment, PLE can only generate elliptically polarized waves rather than orthogonally polarized modes. By explicitly identifying the small $k_{\perp}$ constraint and the associated control parameter $\Delta$, we show that efficient conversion can occur at a limited number of sites in the magnetosphere.  Predictions of the detailed conversion properties will require combining linear mode conversion theory with specific emission mechanisms and models of magnetospheric wave refraction \cite{beskin2012mean} that we leave to future work.

The linear polarization fraction in pulsar radio emission is almost constant below a critical frequency. It decreases at higher frequencies \cite{morris1981depolarisation}, where orthogonal PA jumps and depolarization are frequently observed \cite{manchester1975observations}. Similar orthogonal PA jumps have recently been reported in FRBs \cite{niu2024sudden}. In our framework, the critical frequency corresponds to the local resonance frequency $\omega_{\rm res}$, below which O--X conversion is suppressed and the signal remains almost linearly polarized. Above $\omega_{\rm res}$, enhanced O--X conversion produces mixed circular and orthogonally polarized components, stochastically distributed across the radio beam. It naturally explains the observed elliptically polarized waves, orthogonal PA jump and depolarization effects.

\begin{acknowledgments}
We thank Victoria Kaspi, Alexander Philippov, Anatoly Spitkovsky, and Lorenza Viola for valuable discussions. D.D. acknowledges the Tsinghua-Princeton exchange program that started this work. A.B. is supported by a PCTS fellowship and a Lyman Spitzer Jr. fellowship. J.F.M. acknowledges support from NSF grant AST-2508744. Simulations were performed on Dartmouth College's \textit{Discovery} cluster.
\end{acknowledgments}

\bibliographystyle{apsrev4-2}
\bibliography{bibliography}

\newpage
\onecolumngrid
\begin{center}
    \textbf{End Matter}
\end{center}

\twocolumngrid
\textit{Appendix A: Wave Dispersion in Ultramagnetized Pair Plasmas}---By taking the Fourier transform of the Vlasov-Maxwell equations, we can obtain the equations for electromagnetic waves propagating in a relativistic plasma,
\begin{align}
\boldsymbol{k} \times(\boldsymbol{k} \times \boldsymbol{E})+\frac{\omega^2}{c^2} \boldsymbol{\epsilon} \cdot \boldsymbol{E}=0\,.
\label{wave equation}
\end{align}
If we assume the background magnetic field as infinitely strong and along the $\boldsymbol z$ direction, $\boldsymbol \epsilon$ can be written as
 \begin{align}
\boldsymbol{\epsilon}=\left(\begin{array}{ccc}
1\ \ & 0& 0 \\
0\ \ & 1& 0 \\
0\ \ & 0& 1-\omega_{\text{p,eff}}^2/\omega^2
\end{array}\right)\,.
\label{dieletric tensor}
\end{align}
where $\omega_{\text{p,eff}}$ is determined for specific plasma momentum distribution. We have $k_{\|}=k_z$, and choosing $\boldsymbol k$ in the $\boldsymbol x-\boldsymbol z$ plane yields $k_x=k_{\perp}, k_y=0$. Equation~(\ref{wave equation}) can then be written as
\begin{equation}
    \begin{pmatrix}
        k_{\parallel}^2 c^2-\omega^2 & 0 & -k_{\perp} k_{\parallel} c^2 \\
        0 & k^2 c^2-\omega^2 & 0 \\
        -k_{\perp} k_{\parallel} c^2 & 0 & k_{\perp}^2 c^2-\omega^2+\omega_{\text{p,eff}}^2
    \end{pmatrix}\cdot \bm{E} = 0\ .
    \label{Matrix}
\end{equation}
The determinant of the matrix gives the dispersion relation in Equation~(\ref{Dispersion relation}).

\begin{figure*}
\includegraphics[width=1\textwidth]{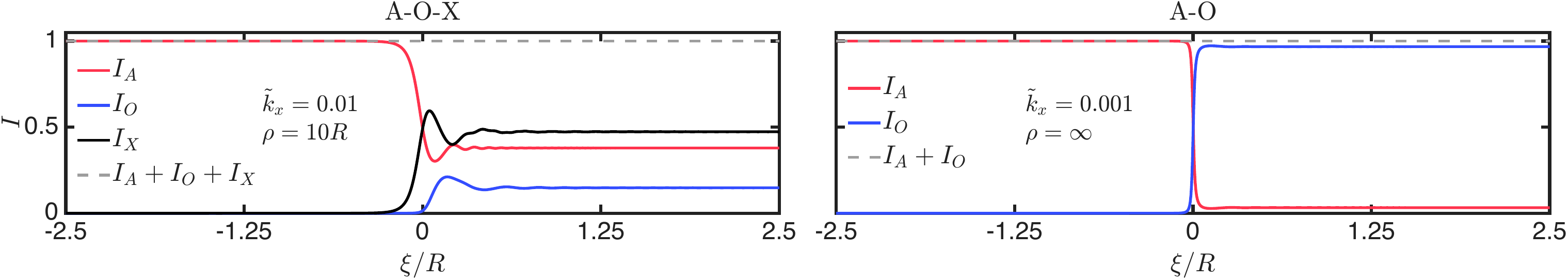}
\vspace{-16pt}
\caption{Mode energy evolution during conversion at the mode-resonance point. Three mode conversion is shown on the left panel considering longitudinal plasma gradient and field line bending effect. A--O conversion is shown on the right panel when field line bending is neglected. Both cases are with parameters $\omega_{\mathrm{p}}=10^8\mathrm{Hz},\ \gamma_0=10,\ L_{\mathrm{p}}=R,\ \varphi=\pi/2$ except different choice of $\tilde{k}_x$ and $\rho$.}
\label{fig:three mode}
\end{figure*}

\textit{Appendix B: Three Mode Conversion at the Resonant Frequency}---Around the mode-resonance point $\omega\approx \omega_{\rm res}$, plasma inhomogeneities become dynamically important in A--O conversion. For simplicity, we assume a linear plasma density gradient in the longitudinal direction. Transverse gradients can be considered through conservation of wave momentum perpendicular to the gradient direction, $\boldsymbol{k\times\nabla}n=\mathrm{const}$. Using $k_{\|}=k_0(1+\lambda)$ and $\omega_\mathrm{p}(\xi)=\omega_\mathrm{p}(1-\xi/L_\mathrm{p})$, the lower right term in Equation~(\ref{dieletric tensor}) should be expanded as
\begin{align}
    \frac{\omega_\mathrm{p}^2}{\gamma_0^3\left(\omega-\beta c k_{\|}\right)^2}=1+2(f\lambda+g\xi)\,,
\label{expansion}
\end{align}
where
\begin{align}
    f=\frac{\beta_0}{1-\beta_0} \quad g=\frac{\omega_\mathrm{p}/L_\mathrm{p}}{\gamma_0^{3/2}(1-\beta_0)ck_0}\,.
\end{align}
Here, $L_\mathrm{p}$ is the length scale of the change in plasma frequency. We substitute Equation~(\ref{expansion}) into Equation~(\ref{wave equation vector}), and follow the same procedure as above to obtain
\begin{align}
	\left(\begin{array}{ccc}
		2\lambda & 0 & \tilde{k}_x-\delta_x \\
		0 & 2\lambda & \delta_y \\
		\tilde{k}_x-\delta_x &  \delta_y & 2(f\lambda+g\xi)
	\end{array}\right)\cdot\left(\begin{array}{c}
		E_{x} \\
		E_{y} \\
		E_{z}
	\end{array}\right)=0\,.
    \label{expanded matrix 2}
\end{align}
Contrary to two-mode conversion, $E_z$ and $E_x,E_y$ are now of the same order, and all matrix terms are preserved to first order. Again, we can obtain the three eigenvalues of the matrix, which correspond to the independent evolution of three modes, with $\lambda_X=0$ and 
\begin{align}
        \lambda_{\rm{O/A}}=\frac{-g\xi\pm\sqrt{g^2\xi^2+f[(k_x-\delta_x)^2+\delta_y^2]}}{2f}\,.
\label{lambda_threemode}
\end{align}
We substitute $k_z$ with $-i{\rm d}/{\rm d}\xi$ to find
\begin{align}
\left(i\frac{{\rm d}}{{\rm d}\xi}+k_0\right)
\begin{pmatrix} E_x \\ E_y \\ fE_z \end{pmatrix}=\frac{k_0}{2}
\begin{pmatrix} 
0 & 0 & \tilde{k}_x-\delta_x \\ 
0 & 0 & \delta_y \\ 
\tilde{k}_x-\delta_x & \delta_y & 2g\xi 
\end{pmatrix}\cdot
\boldsymbol{E}\,.
\label{ODE_3mode}
\end{align}
For many parameter choices, the ODEs in Equation~(\ref{ODE_3mode}) can only be solved numerically. Figure \ref{fig:three mode} shows the three mode energy flux evolution and energy flux conservation. The energy flux of each modes are calculated by
\begin{align}
    I^i=c|E_x^i|^2+c|E_y^i|^2+cf|E_z^i|^2\,,
\end{align}
where $i=\rm A,O,X$. Term $cf|E_z|^2$ can be understood as an acoustic flux \cite{stix1992waves} associated with the longitudinal particle motion.

For $\theta=0$, we neglect magnetic field line bending. Then, X-modes are decoupled from A- and O-modes, and energy transfer only happens in A--O conversion due to longitudinal plasma gradient effects (see Figure \ref{fig:three mode}, right panel). Analytic solutions to Equation~(\ref{ODE_3mode}) are obtained in parabolic cylinder functions. Asymptotically, the conversion efficiency is 
\begin{align}
    \eta_{\rm A\rightarrow O}=\exp{\left(\frac{-\pi k_x^2L_p}{2k_0}\right)}\,.
\end{align}
which is the classic LZ result \cite{landau1932theory,zener1932non}, generalized linear mode conversion theory gives similar result \cite{PhysRevLett.70.1799}. Note that in A--O conversion, only one mode crossing point exist (defined as $\xi_c$ that $\lambda_O(\xi_c)=\lambda_A(\xi_c)$, Equation~\ref{lambda_threemode}), while for A--X and O--X conversion two mode crossing points are degenerate on the real axis as complex conjugates in the complex plane ($\lambda_{\rm O/A}(\xi_c)=\lambda_{\rm X}(\xi_c)$, Equation~\ref{eigenvalue}). This accounts for the $(1-p)$ factor on top of the LZ formula (Equation~\ref{fit}), which gives inefficient conversion for $\Delta\rightarrow0$. The difference can also be seen by comparing Figure~\ref{fig:three mode} right panel and Figure~\ref{fig:Evolution}.

\textit{Appendix C: Asymptotic Estimates of the Conversion Efficiency}---The coupled, first-order ODEs of wave amplitudes (Equation~\ref{ODE}) can be compared with the time-dependent Shrödinger equation of a two-state process in quantum mechanics. We analyze the Hamiltonian matrix
\begin{align}
    i \hbar \frac{{\rm d}}{{\rm d} t} \boldsymbol{\psi}(t)=\left(\begin{array}{cc}
\epsilon(t) & V(t) \\
V(t) & -\epsilon(t)
\end{array}\right)\cdot\boldsymbol{\psi}(t)\,,
\end{align}
where $\boldsymbol{\psi}(t)$ are the probability amplitudes on the diabatic basis states. The function $\epsilon(t)$ is the diabatic energy level and $V(t)$ is the diabatic coupling. We can expand the solution of $\boldsymbol{\psi}(t)$ on the adiabatic basis as
\begin{align}
\boldsymbol{\psi}(t)&= a_1(t) \exp \left[-i \int_0^t {\rm d} \tau\, E_1(\tau) / \hbar\right] \boldsymbol{\chi}_1(t) \\ &+a_2(t) \exp \left[-i \int^t_0 {\rm d} \tau\, E_2(\tau) / \hbar\right] \boldsymbol{\chi}_2(t)\,,
\end{align}
where the adiabatic energy levels $E_{1,2}(t)=\pm\sqrt{\epsilon(t)^2+V(t)^2}$ are the eigenvalues of the Hamiltonian, and $\boldsymbol{\chi}_{1,2}(t)$ are the corresponding eigenvectors. The amplitude evolution of $a_{1,2}(t)$ is determined by
\begin{align}
\begin{split}
        a_2'&=\gamma\exp\left[i\Delta(t)/\hbar\right]a_1\,,\\a_1'&=-\gamma\exp\left[-i\Delta(t)/\hbar\right]a_2\,.
\end{split}
\label{adiabatic}
\end{align}
where $\gamma$ is the adiabatic coupling, given by
\begin{align}
    \gamma=\frac{\dot{\epsilon}(t)V(t)-\epsilon(t)\dot{V}(t)}{2(\epsilon(t)^2+V(t)^2)}
\end{align}
and 
\begin{align}
    \Delta(t)=\int_0^t\left[E_1(t)-E_2(\tau)\right]{\mathrm d}\tau
    \label{Delta}
\end{align}
Integration of Equation~(\ref{adiabatic}) for $t\in\left(-\infty,\infty\right)$ predicts the asymptotic solution. \citet{davis1975nonadiabatic} show that the integration along the real axis can be deformed to a contour $\mathrm{Im}\left[\Delta(t)\right]=\mathrm{Im}\left[\Delta(t_c)\right]$ in the complex plane, with a semicircle detour around $t_c$, where $t_c$ is the zero point of $E_1(t)-E_2(t)$ in the complex plane. We define
\begin{align}
\tilde{a}_2(t)=\exp \left[-i \Delta(t_c) / \hbar\right] a_2(t)\,, \quad \tilde{a}_1(t)=a_1(t)\,,
\end{align}
and rewrite Equation~(\ref{adiabatic}) as
\begin{align}
\begin{split}
\tilde{a}_2^{\prime}&=\gamma \exp \left[i\left(\Delta-\Delta(t_c)\right) / \hbar\right] \tilde{a}_1\,,\\ \tilde{a}_1^{\prime}&=-\gamma \exp \left[-i\left(\Delta-\Delta(t_c)\right) / \hbar\right] \tilde{a}_2\,.
\end{split}
\end{align}
With initial conditions $\tilde{a}_1(-\infty)=1$ and $\tilde{a}_2(-\infty)=0$. Further analysis in the adiabatic limit($\hbar\rightarrow0$), shows that the asymptotic solution is $\tilde{a}_2(\infty)=1$ and $a_2(\infty)=\exp \left[-i \Delta(t_c) / \hbar\right]$. The transition probability is
\begin{align}
    p=\exp[-\frac{2\mathrm{Im}\left[\Delta(t_c)\right]}{\hbar}].
\end{align}
Direct comparison with Equation~(\ref{ODE}) shows that $\hbar=1$, $t=\xi$. $\epsilon(\xi)$ is half the difference between diagonal terms in the Hamiltonian, and $V(\xi)$ is the off-diagonal term:
\begin{align}
    \epsilon=\frac{\delta_y^2-(\tilde{k}_x/\varepsilon-\delta_x)^2}{4(1-1/\varepsilon^2)}, \ \ \ V=\frac{\delta_y(\tilde{k}_x/\varepsilon-\delta_x)}{2(1-1/\varepsilon^2)}\,.
\end{align}
The transformation of the diabatic basis to an adiabatic basis is simply the rotation between our $(\boldsymbol{x},\boldsymbol{y},\boldsymbol{z})$ coordinate system and A--O--X coordinates, the difference between $E_{\pm}$ is the difference between $\lambda_O$ and $\lambda_X$:
\begin{align}
    E_{\pm}=\pm\frac{\delta_y^2+(\tilde{k}_x/\varepsilon-\delta_x)^2}{4(1-1/\varepsilon^2)}
\end{align}
Zero-points for $E_{+}(\xi_c)-E_{-}(\xi_c)=0$ are on the complex plane, $\xi_c=\tilde{k}_x\rho(\cos\varphi\pm i\sin\varphi)$, which is degenerate on the $\boldsymbol{x}$-axis. With Equation~\ref{Delta}, we find
\begin{align}
    \Delta(\xi_c)=\int_0^{\xi_c}(E_{+}-E_{-})d\xi=\frac{i\tilde{k}_x^3\rho\sin^3\varphi}{3|\epsilon^2-1|}\,.
\end{align}
Further analysis following the DDP method \cite{davis1975nonadiabatic} yields the $\sqrt{2}$ coefficient in the DDP formula used above. We notice that rapid conversion always happens within a narrow region around the conversion point. The location $r$ of the conversion point and the length scale $l$ of the rapid-conversion region are
\begin{align}
    r\sim {\rm Re}\left(\xi_c\right)\sim\tilde{k}_x\rho\cos\varphi\,,\ \ l \sim {\rm Im}\left(\xi_c\right)\sim\tilde{k}_x\rho\sin\varphi\,.
\end{align}
We understand $l$ as the radius of convergence for the singularity $\xi_c$ in the complex plane \cite{PhysRevA.59.988}. Asymptotic approximation and linear analysis hold simultaneously when $l\ll L\ll\rho$, where $L$ is the system length scale. To achieve linear conversion as discussed in this Letter, we need $L$ to be well-defined, which is ensured if $\tilde{k}_x\ll1$. Equation~(\ref{condition}) shows that this is true for most cases. 

\end{document}